\documentclass[12pt,a4paper]{article}
\usepackage{graphicx}
\def\Journal#1#2#3#4{{#1} {\bf #2} (#3) #4}

\def\PLB{{\rm Phys. Lett.}  B}
\def\PRL{\rm Phys. Rev. Lett.}
\def\PRD{{\rm Phys. Rev.} D}

\begin{document}
\begin{center}
{\large Determination of the $\psi(3770)$, $\psi(4040)$,
$\psi(4160)$ and $\psi(4415)$}\\
{\large resonance parameters}\\
\end{center}

{\small{ M.~Ablikim$^{1}$, J.~Z.~Bai$^{1}$, Y.~Ban$^{12}$,
X.~Cai$^{1}$, H.~F.~Chen$^{17}$, H.~S.~Chen$^{1}$, H.~X.~Chen$^{1}$,
J.~C.~Chen$^{1}$, Jin~Chen$^{1}$, Y.~B.~Chen$^{1}$
% ...
  \\~~~~~~~~~~~~~~~~~~~~~~~~~~~~~~~~~~~~~~~~~~~(BES Collaboration) \\
\vspace{0.2cm}
$^{1}$ Institute of High Energy Physics, Beijing 100049, People's Republic of China\\
% ...
$^{12}$ Peking University, Beijing 100871, People's Republic of China\\
% ...
$^{17}$ University of Science and Technology of China, Hefei 230026, People's Republic of China\\
% ...
}

\date{\today}
\graphicspath{{figure/}}

\begin{abstract}
$R$ measurement data taken with the BESII detector at center-of-mass
energies between 3.7 and 5.0 GeV is fitted to determine resonance
parameters (mass, total width, electron width) of the high mass
charmonium states, $\psi(3770)$, $\psi(4040)$, $\psi(4160)$ and
$\psi(4415)$. Various effects, including the relative phases between
the resonances, interferences, the energy-dependence of the full
widths, and the initial state radiative correction, are examined.
The results are compared to previous studies.
\end{abstract}

\section{Introduction}

The total cross section for hadron production in $e^+e^-$
annihilation is usually parameterized in terms of the ratio $R$,
which is defined as $R=\sigma(e^+e^- \rightarrow \mbox{hadrons})/
\sigma(e^+e^-\rightarrow \mu^+\mu^-)$, where the denominator is the
lowest-order QED cross section, $\sigma
(e^+e^-\rightarrow\mu^+\mu^-)=\sigma^0_{\mu \mu}=4\pi\alpha^2 / 3s$.
The measured $R$ values are consistent with the three-color quark
model predictions. At the open flavor thresholds where resonance
structures show up, measurements of $R$ value are used to determine
resonance parameters. For the high mass charmonium resonances, the
$\psi(3770)$ was measured by MARK-I \cite{MarkI3770}, DELCO
\cite{DELCO}, MARK-II \cite{MarkII} and BES
\cite{bes3770}\cite{besnondd}; the $\psi(4040)$ and $\psi(4160)$
were measured by DASP~\cite{DASP}; and the $\psi(4415)$ was measured
by DASP \cite{DASP} and MARK-I \cite{MarkI4415}. There were also
some other measurements of $R$ values as reported in
Refs.~\cite{MARKIwide,PLUTO,CB}, but no attempt was made to
determine resonance parameters. The resonance parameters in the
Particle Data Group~(PDG)'s compilation remained unchanged for more
than 20 years up to the 2004 edition \cite{pdg04}. The resonance
parameters for the three high mass resonances were updated by
PDG2006 \cite{pdg06}, based on K. Seth's evaluation \cite{seth}
using combined BESII \cite{besr99} and Crystal Ball \cite{CB} data.

The most recent and precise $R$ measurements between 2-5 GeV were
made by BESII~\cite{besr99}. Experimentally, $R$ for both the
continuum and the wide resonance region is given by
\begin{equation}
R_{exp}=\frac{ N^{obs}_{had} - N_{bg}} { \sigma^0_{\mu\mu}  L
\epsilon_{trg} \epsilon_{had}(1+\delta_{obs})}, \label{rexp}
\end{equation}
where $N^{obs}_{had}$ is the number of observed hadronic events,
$N_{bg}$ is the number of the residual background events, $L$ is the
integrated luminosity, $(1+\delta_{obs})$ is the effective
correction factor of the initial state radiation
(ISR)~\cite{CB}\cite{isrhu}, $\epsilon_{had}$ is the detection
efficiency for hadronic events determined by the Monte Carlo
simulation without bremsstrahlung being simulated, and
$\epsilon_{trg}$ is the trigger efficiency. The determination of $R$
values and resonance parameters are intertwined; the factor
$(1+\delta_{obs})$ in Eq.~(\ref{rexp}) contains contributions from
the resonances and depends on the resonance parameters. Therefore,
the procedure to calculate $(1+\delta_{obs})$ requires a number of
iterations before stable results can be obtained.

In this work, we perform a global fit over the entire center-of-mass
energy region from 3.7 to 5.0 GeV covering the four resonances,
$\psi(3770)$, $\psi(4040)$, $\psi(4160)$ and $\psi(4415)$, and
include interference effects among the resonances. We also adopt
energy-dependent full widths, and introduce relative phases between
the resonances.  Finally, the new $R$ values due to the updated
resonant parameters are compared to the published ones
\cite{besr99}.

\section{Fitting of the resonant parameters}

The phenomenological models and formulas used in our fitting are
briefly described below.

\subsection{Breit-Wigner form}

The relativistic Breit-Wigner amplitude for $e^+e^-$ $\to$ resonance
$\to$ final state $f$ is
\begin{equation}
{\cal T}_r^{f}(W)
=\frac{M_r\sqrt{\Gamma_r^{ee}\Gamma_r^f}}{W^2-M_r^2+ iM_r\Gamma_r}
e^{i\delta_r}, \label{rlaf}
\end{equation}
where $W\equiv E_{cm}\equiv\sqrt{s}$ is the center-of-mass energy,
the index $r$ represents the resonance to be studied, $M_r$ is the
nominal mass, $\Gamma_r$ is the full width, $\Gamma_r^{ee}$ is the
electron width, $\Gamma_r^{f}$ is the hadronic width for the
decaying channel $f$, and $\delta_r$ is the phase.

The high mass charmonia decay into several two-body final states
$f$. According to the Eichten model~\cite{eichten} and experimental
data~\cite{barnes}, the decay channels (including their conjugate
states) are:
\begin{eqnarray}
\psi (3770)&\Rightarrow& D\bar{D};\nonumber\\
\psi (4040)&\Rightarrow& D\bar{D},D^{\ast}\bar{D}^{\ast},
                         D\bar{D}^{\ast},
                         D_s\bar{D}_s;\nonumber\\
\psi (4140)&\Rightarrow& D\bar{D},D^{\ast}\bar{D}^{\ast},
                         D\bar{D}^{\ast},
                         D_s\bar{D}_s,D_s\bar{D}_s^{\ast};\nonumber\\
\psi (4415)&\Rightarrow& D\bar{D},D^{\ast}\bar{D}^{\ast},
                         D\bar{D}^{\ast},
                         D_s\bar{D}_s,D_s\bar{D}_s^{\ast},D_s^{\ast}\bar{D}_s^{\ast}, D\bar{D_{1}}, D\bar{D^{*}_{2}}.\nonumber
\end{eqnarray}

The total squared inclusive amplitude of the resonances is the
incoherent sum over all different decay channels $f$,
\begin{equation}
|{\cal T}_{res}|^2=\sum_{f} |\sum_{r} {\cal T}_{r}^{f}(W)|^2.
\end{equation}
The resonant cross section expressed as the $R$ value is then given
by
\begin{equation}
R_{res} = \frac{\sigma_{res}}{\sigma_{\mu\mu}^{0}} =\frac{12\pi}{s}
|{\cal T}_{res}|^2. \label{eqres}
\end{equation}

\subsection{Energy-dependence of the full width}

The full width of a broad resonance depends on the energy.
A phenomenological model derived from quantum mechanics is used to
describe the behavior of $\Gamma_r^{f}(W)$,
which depends on the momentum and the orbital angular momentum $L$
of the decaying final state~\cite{blatt},
\begin{equation}
\Gamma_{r}^{f}(W)=\hat{\Gamma}_r\frac{2M_r}{M_r+W}\sum\limits_L
\frac{Z_{f}^{2L+1}}{B_L},\label{widtqm}
\end{equation}
where, $\hat{\Gamma}_r$ is a parameter to be determined by fitting
experimental data, $Z_{f}\equiv \rho P_{f}$, $\rho$ is the radius of
the interaction with the order of a few fermis (the value is
insensitive to the physical results), $P_{f}$ is the decay momentum,
and $2M_r/(M_r+W)$ is a relativistic correction factor. The
energy-dependent partial wave functions $B_L$ are given in
Ref.~\cite{blatt}:
\begin{displaymath}
B_0=1,~~B_1=1+Z^2,~~B_2=9+3Z^2+Z^4,
\end{displaymath}
\begin{equation}
B_3=225+45Z^2+6Z^4+Z^6.
\end{equation}

When the resonance decays to several hadronic channels, the total
hadronic width is the sum of all its partial widths,
\begin{equation}
\Gamma_r^{had}(W)=\sum_{f}\Gamma_{r}^{f}(W).
\end{equation}
The total width of the resonance $r$ is expressed as
\begin{equation}\label{widthtotal}
\Gamma_r(W)=3\Gamma_r^{ee}+\Gamma_r^{had}(W),
\end{equation}
where the universality of the leptons is used, i.e.
$\Gamma_r^{ee}=\Gamma_r^{\mu\mu}=\Gamma_r^{\tau\tau}$, and
$\Gamma_r^{ee}$ is the experimental electron width, which includes
the contribution from the vacuum polarization effect.

\subsection{Continuum background}

The contribution of continuum background originating from initial
light quark pairs $u\bar{u}$, $d\bar{d}$ and $s\bar{s}$, which is
written as $R_{QCD}^{(uds)}$, may be predicted by pQCD above 2
GeV\cite{pdg06}. Being close to the production threshold, the
continuum open-$c\bar{c}$ background can only be described by
phenomenological models or empirical expressions. Since there are
many possible channels above the open-charm threshold, and their
cross sections are expected to vary smoothly, we parametrize, for
simplicity, the continuum charm background with a second order
polynomial,
\begin{equation}
R_{con}^{(c)} = C_0 + C_1 (W - 2M_{D^\pm}) + C_2 (W - 2M_{D^\pm})^2,
\label{poly}
\end{equation}
where $C_0$, $C_1$ and $C_2$ are free parameters, and $M_{D^\pm}$ is
the mass of the lightest meson $D^\pm$.

\subsection{Fitting scheme}

We fit the experimental data with MINUIT~\cite{minuit} using a least
squares method, with $\chi^2$ defined as~\cite{chiequ}
\begin{equation}
\chi^2 = \sum\limits_i \frac{[f_c\widetilde{R}_{exp}(W_i) -
\widetilde{R}_{the}(W_i)]^2} {[f_c\Delta
\widetilde{R}^{(i)}_{exp}]^2}+\frac{(f_c-1)^2}{\sigma^{2}_{c}},
\label{fitf2}
\end{equation}
where $W_i$ stands for the energy of the measured point. The
experimental quantity
\begin{equation}
\widetilde{R}_{exp}=\frac{ N^{obs}_{had} - N_{bg} } {
\sigma^0_{\mu\mu}  L  \epsilon_{trg} \epsilon_{had}}, \label{rlike}
\end{equation}
and the corresponding theoretical quantity
\begin{equation}
\widetilde{R}_{the}=(1+\delta_{obs}) R_{the}, \label{fitr}
\end{equation}
where,
\begin{equation}
R_{the}=R_{con}+R_{res},\label{rthe}
\end{equation}
and
\begin{equation}
R_{con}=R^{(uds)}_{(QCD)}+R_{con}^{(c)}.
\end{equation}
$\Delta \widetilde{R}^{(i)}_{exp}$ in Eq.~(\ref{fitf2}) is the
combined statistical and uncommon systematic errors of
$\widetilde{R}_{exp}(W_i)$; the error common to all the points
$\sigma_{c}~(\sim 3.3\%)$ is not included. In each iteration, the
resonant parameters in the calculation of $(1+\delta_{obs})$ and
$R_{the}$ are updated to new values. $f_c$ is a scale factor which
reflects the influence of the common error on the fitting.
$R^{(uds)}_{(QCD)}$ is derived from pQCD~\cite{rpqcd}.

The free parameters in the fit are $M_r$, $\Gamma_r^{ee}$,
$\delta_r$ in Eq.~(\ref{rlaf}), $\hat{\Gamma}_r$ in
Eq.(\ref{widtqm}), and $C_0$, $C_1$, $C_3$ in Eq.~(\ref{poly}). Only
relative values of the phases can be extracted, so for simplicity,
the phase of $\psi(3770)$ is set to zero.

\section{Results and discussion}

The resonant parameters of the high mass charmonia determined in
this work, together with those in PDG2004, PDG2006 and the results
given in Ref.~\cite{seth} are listed in Table~\ref{table}. The
updated $R$ values between 3.7 and 5.0 GeV and the fit curves are
illustrated in Fig.~\ref{resbes}. The fit yields
$\chi^2/d.o.f=1.06$, which indicates a reasonable fit.

In order to understand the uncertainty of the model-dependence,
alternative choices and combinations of Breit-Wigner forms,
energy dependence of the full width predicted by quantum mechanics
model~\cite{blatt} or the effective interaction
theory~\cite{gaocsg}, and continuum charm background described by a
second order polynomial or the phenomenological form used by
DASP~\cite{DASP} are used. We find that the results are sensitive to
the form of the energy dependent total width, but not sensitive to
the form of background. The DASP background function has six
continuum production channels, while the effective interaction
theory predicts different energy-dependence of the hadronic width
for different decay channels. However in both cases the best fits
give unreasonable results. This may be understood as the inclusive
data can not supply enough information to determine the relative
strength of different decay channels. To understand the detailed
structures and components of the high mass charmonia states, it is
necessary to collect data at each peak with sufficiently high
statistics, and to develop more reliable physical models. This is
one of the physics tasks of a tau charm factory, and might be
further studied with BESIII that is under construction.

It is worth noting that the change of the resonance parameters
affects the effective initial state radiative correction factors,
and thus affects the $R$ values. Fig.~\ref{dr} shows comparison
between the $R$ values published in~\cite{besr99} and the updated
values in this work; the difference varies with the resonant
structure. In general the relative difference is within $3\%$, and
for a few energy points the maximum difference is about $6\%$. Our
results are in agreement with the previous experiments in most
cases, but large differences are observed in some of the parameters,
such as the mass of the $\psi(4160)$. This is mainly due to the
reconsideration of the radiative correction factors, and the
inclusion of interferences between the resonances.


\begin{table*}[htbp]
\caption{The resonance parameters of the high mass charmonia in this
work together with the values in PDG2004~\cite{pdg04},
PDG2006~\cite{pdg06} and K. Seth's evaluations~\cite{seth} based on
Crystal Ball and BES data. The total width $\Gamma_{tot}\equiv
\Gamma_r(M)$ in Eq.(\ref{widthtotal}).} \vskip -0.1cm \hskip 1.5cm
\parbox[t]{12cm}{
\begin{center}
\begin{tabular}{|c|c|c|c|c|c|} \hline \hline
        &         &$\psi(3770)$  &$\psi(4040)$&$\psi(4160)$&$\psi(4415)$\\ \hline
        &PDG2004  &3769.9$\pm$2.5&4040$\pm$10 &4159$\pm$20 & 4415$\pm$6 \\
        &PDG2006  &3771.1$\pm$2.4&4039$\pm$1.0&4153$\pm$3  & 4421$\pm$4 \\
$M$     &CB (Seth) &   -          &4037$\pm$2  &4151$\pm$4  & 4425$\pm$6 \\
(MeV/$c^2$)   &BES (Seth)&   -          &4040$\pm$1  &4155$\pm$5  & 4455$\pm$6 \\
        &BES (this work) &3771.4$\pm$1.8& 4038.5$\pm$4.6 & 4191.6$\pm$6.0 & 4415.2$\pm$7.5\\\hline\hline

              &PDG2004  &23.6$\pm$2.7&52$\pm$10    &78$\pm$20   &     43$\pm$15   \\
              &PDG2006  &23.0$\pm$2.7&80$\pm$10    &103$\pm$8    &     62$\pm$20   \\
$\Gamma_{tot}$&CB (Seth) &  -        &85$\pm$10     &107$\pm$10&119$\pm$16 \\
(MeV)         &BES (Seth)&  -        &89$\pm$6     &107$\pm$16&118$\pm$35\\
              &BES (this work)      &25.4$\pm$6.5&81.2$\pm$14.4&72.7$\pm$15.1&73.3$\pm$21.2\\\hline\hline

             &PDG2004  &0.26$\pm$0.04&0.75$\pm$0.15&0.77$\pm$0.23&0.47$\pm$0.10 \\
             &PDG2006  &0.24$\pm$0.03&0.86$\pm$0.08&0.83$\pm$0.07&   0.58$\pm$0.07 \\
$\Gamma_{ee}$&CB (Seth) &   -         &0.88$\pm$0.11&0.83$\pm$0.08&0.72$\pm$0.11\\
(keV)        &BES (Seth)&   -         &0.91$\pm$0.13&0.84$\pm$0.13&0.64$\pm$0.23\\
             &BES (this work)      &0.18$\pm$0.04&0.81$\pm$0.20&0.50$\pm$0.27&0.37$\pm$0.14\\\hline\hline
$\delta$ (degree)&BES (this
work)&0&133$\pm$68&301$\pm$61&246$\pm$86\\\hline\hline
\end{tabular}
\label{table}
\end{center}}
\end{table*}

\begin{center}
\begin{figure*}
\includegraphics[width=12cm,height=9cm]{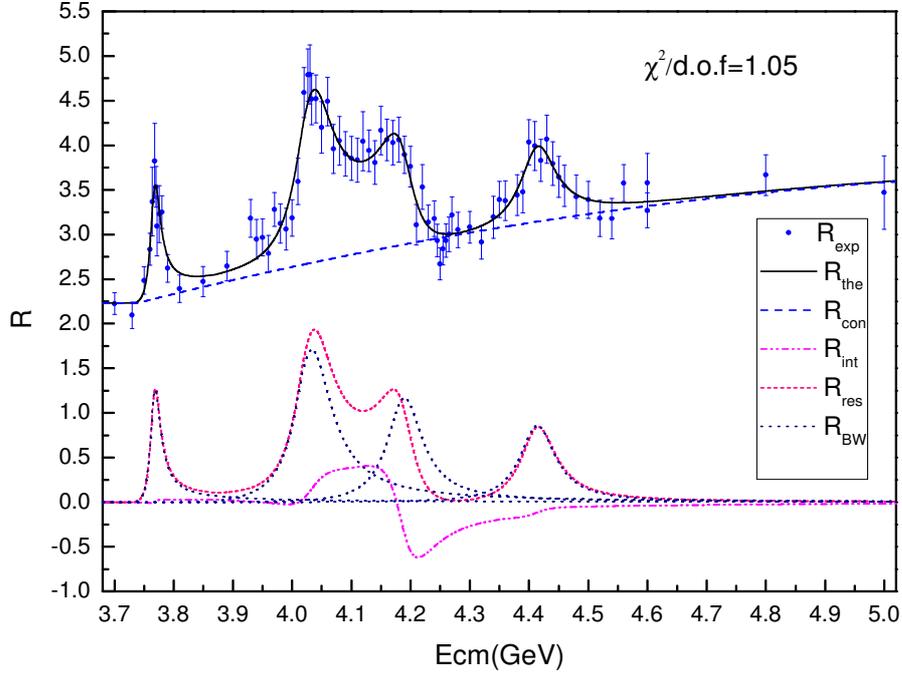}
\vspace{-5mm}\caption{The fit to the $R$ values for the high mass
charmonia structure. The dots with error bars are the updated $R$
values. The solid curve shows the best fit, and the other curves
show the contributions from each resonance $R_{BW}$,  the
interference $R_{int}$, the summation of the four resonances
$R_{res}=R_{BW} + R_{int}$, and the continuum background $R_{con}$
respectively.} \label{resbes}
\end{figure*}
\end{center}

\begin{center}
\begin{figure*}
\includegraphics[width=8cm,height=6cm]{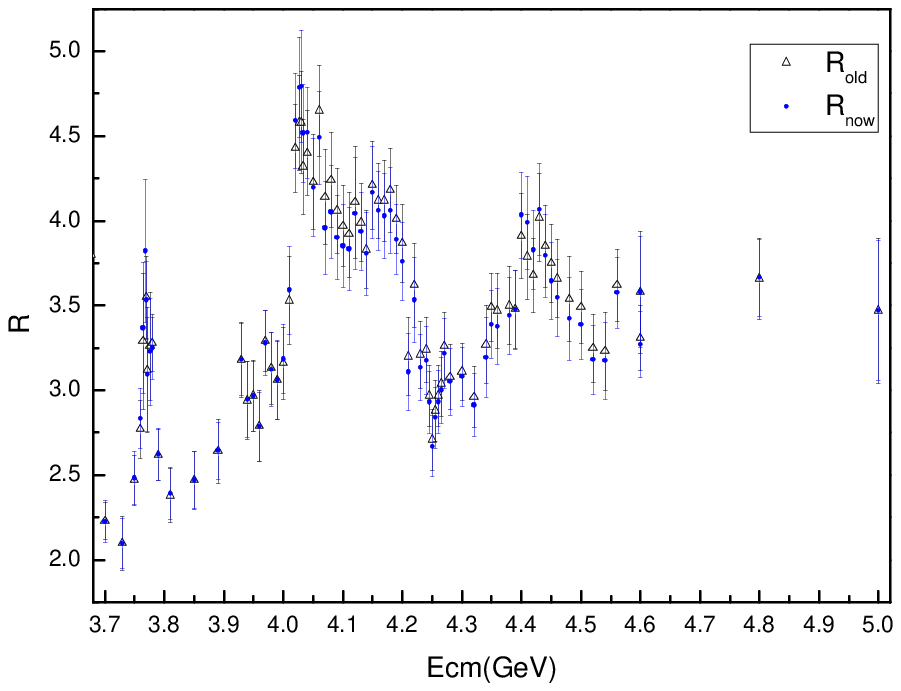}
\includegraphics[width=8cm,height=6.1cm]{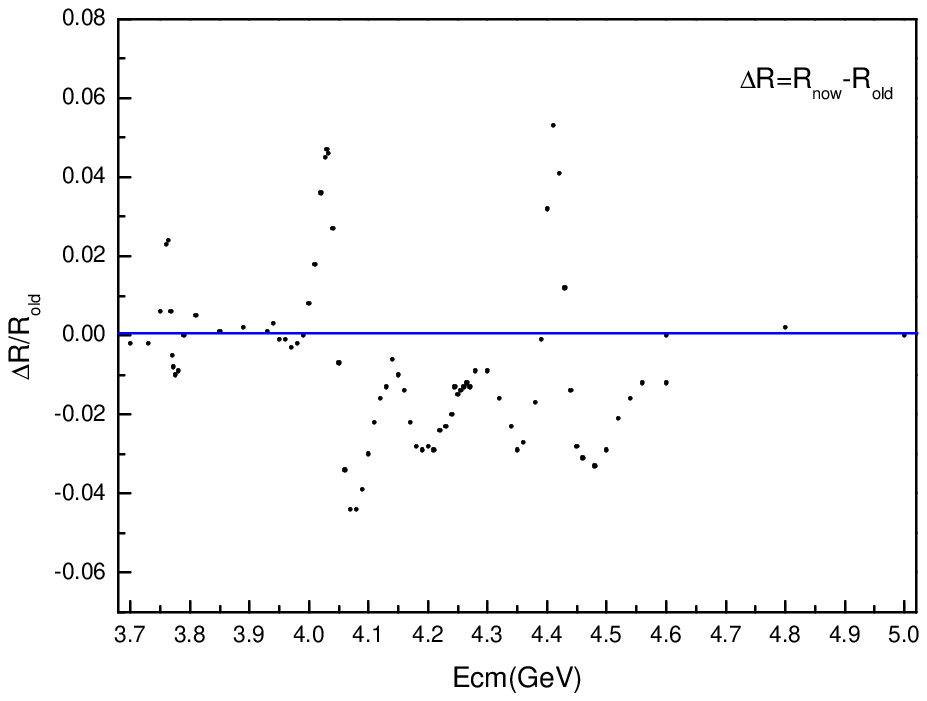}
\put(-345,150){\bf \large (I)} \put(-115,150){\bf \large (II)}
\vspace{-5mm}\caption{(I) The comparison of $R$ values between the
values published in Ref.~\cite{besr99} (triangles: $R_{old}$) and
the updated values in this work (points: $R_{now}$). (II) The
relative differences between the two sets of $R$ values.} \label{dr}
\end{figure*}
\end{center}
\vspace{-1.25cm} The BES collaboration thanks the staff of BEPC and
computing center for their hard efforts. This work is supported in
part by the National Natural Science Foundation of China under
contracts Nos. No.19991480, No.19805009, No.19825116, 10491300,
10225524, 10225525, 10425523, the Chinese Academy of Sciences under
contract No. KJ 95T-03, the 100 Talents Program of CAS under
Contract Nos. U-11, U-24, U-25, and the Knowledge Innovation Project
of CAS under Contract Nos. U-602, U-34 (IHEP), the National Natural
Science Foundation of China under Contract No. 10225522 (Tsinghua
University), and the Department of Energy under Contract
No.DE-FG02-04ER41291 (U. Hawaii).

\end{document}